\documentclass[aps,prb,twocolumn,reprint,superscriptaddress]{revtex4-1}
\usepackage{amsmath,amssymb}
\usepackage{bm}
\usepackage{graphicx}
\usepackage{xcolor}

\begin{document}

\title{Pseudo-drag of a polariton superfluid}

\author{Igor Y. Chestnov}

\affiliation{Westlake University, School of Science, 18 Shilongshan Road, Hangzhou 310024, Zhejiang Province, China}
\affiliation{Westlake Institute for Advanced Study, Institute of Natural Sciences, 18 Shilongshan Road, Hangzhou 310024, Zhejiang Province, China}
\affiliation{Vladimir State University, Gorkii St. 87, 600000, Vladimir, Russia}

\author{Yuri G. Rubo}
\affiliation{Instituto de Energ\'{\i}as Renovables, Universidad Nacional Aut\'onoma de M\'exico, Temixco, Morelos, 62580, Mexico}

\author{Alexey V. Kavokin}
\affiliation{Westlake University, School of Science, 18 Shilongshan Road, Hangzhou 310024, Zhejiang Province, China}
\affiliation{Westlake Institute for Advanced Study, Institute of Natural Sciences, 18 Shilongshan Road, Hangzhou 310024, Zhejiang Province, China}

\begin{abstract}
The drag of half-light half-mater quasiparticles, exciton-polaritons, by an electric current is a peculiar mechanism of light-matter interaction in solids.
While an ideal superfluid is protected from being dragged by its zero viscosity, here we argue that the state of the superfluid polariton condensate formed by a non-resonant optical pumping can be controlled by the electric current. The proposed mechanism is based on the stimulated relaxation of moving uncondensed excitons dragged by the electric current.
The stimulated relaxation process favors the formation of a moving condensate in a quantum state that is characterised by the lowest condensation threshold. We also show that the electron-mediated inelastic scattering of the reservoir excitons to the condensate leads to the transfer of a non-zero mean momentum to the  electron gas thus contributing to the electric current.
We predict the generation of circular electric currents in a micropillar cavity in the presence of a nonresonant laser pumping at normal incidence.
\end{abstract}



\maketitle

\section{Introduction}

A superfluid state of matter is characterized by zero viscosity, hence it is perfectly protected from being perturbed by a weak external force.  This protection constitutes one of the main experimental signatures of conventional superfluids such as liquid helium \cite{Kapitza1938,Pitaevskii2016}. However, this paradigm needs to be carefully reconsidered if applied to driven-dissipative superfluid systems. In particular, optically pumped bosonic condensates of exciton-polaritons (or simply polaritons) in semiconductor microcavities represent such nonequilibrium systems, where several phenomena consistent with superfluidity have been experimentally demonstrated. Polaritons are hybrid quasiparticles arising due to the strong coupling between excitons and photons.  Polaritons obey the bosonic statistics and they may undergo a transition into the condensed phase. In this phase, the polariton fluid demonstrates a dissipationless propagation with a subsonic velocity through a weak defect, that manifests its superfluid behaviour \cite{Amo2009,Wouters2010_1}.

One would naturally expect that  in the subsonic regime the polariton superfluid  should not be perturbed by an electric current flowing either in the same quantum well or in the neighbouring conducting layer. On the other hand, the coupling is possible for non-condensed polaritons which do not belong to the superfluid. Recently, Berman and co-authors \cite{berman_10} predicted the existence of the mutual drag between the normal fraction of a polariton gas and the electric current. Such drag effect is mediated by the long-range interaction between the excitonic component of polaritons and  charge carriers, leading to the appearance of the flow of the normal polariton fraction induced by the electric current and vice versa.

The indications of a drag of a polariton superfluid by a current have been reported in the recent experimental work \cite{Snoke2018} demonstrating that the speed of a superfluid polariton flow is sensitive to the magnitude and the direction of the electric current flowing in the same quantum well.
Although the particular mechanism governing the detected  effect remains unclear, the observed phenomenon indicates that the interaction between the polaritons and the carriers is not limited only to the conventional Coulomb drag of the normal fraction.

In this paper, we show theoretically how an electric current can affect the propagation of a polariton superfluid. The mechanism that governs this effect is related to the usual drag of the normal excitonic component. We demonstrate that the flow of non-condensed excitons (exciton-polaritons) leads to formation of the condensate with a non-zero momentum.

It is well-known that a driven-dissipative polariton condensate, which is formed by the nonresonant pumping, does not necessary occupy the lowest energy stationary state. In fact, the state of the condensate is determined by the balance between the gain and the losses that defines the threshold of polariton lasing \cite{Aleiner2012}.
If the exciton reservoir moves, the gain  is changed with respect to the static case, and the condensate may be formed in a moving state as well.

In particular, if the exciton reservoir is dragged by the electric current, the wave-vector of the forming condensate will depend on the direction and strength of the current. This effect is phenomenologically equivalent to the drag of a superfluid. However, the predicted phenomenon is not a direct drag effect, strictly speaking. It is mediated by the excitonic reservoir. That is why we shall refer to it as a pseudo-drag effect.


\section{The model system}

To be specific, we consider an optical microcavity containing both an un-doped quantum well and a conducting  layer which confines a free electron gas. The polariton condensate is created in the  microcavity by   non-resonant laser pumping. This kind of pumping implies   excitation of high-momentum excitons which then relax in energy feeding the condensate at the bottom of the lower polariton branch \cite{KavokinMicrocavities}.

We start with the semiclassical kinetic equations \cite{KavokinMicrocavities} for the polariton occupation $N_{\mathbf{k}}$  of the quantum state characterised by a wave vector $\mathbf{k}$:
\begin{equation}\label{Eq:Kinetic}
\partial_{t} N_{\mathbf{k}} = P_{\mathbf{k}} - \left(\Gamma_{\mathbf{k}} + W^{\rm out}_{\mathbf{k}} \right) N_{\mathbf{k}} + W^{\rm in}_{\mathbf{k}}\left(N_{\mathbf{k}} +1 \right),
\end{equation}
where $P_{\mathbf{k}}$ describes the generation of  particles in the state $\mathbf{k}$ by the incoherent pumping,
$\Gamma_{\mathbf{k}}$ is the decay rate to the exterior of the microcavity,
$W^\mathrm{out}_\mathbf{k}
=\sum_{\mathbf{k}^\prime}\mathcal{W}_{\mathbf{k}\rightarrow\mathbf{k}^\prime}\left(N_{\mathbf{k}^\prime}+1\right)$
and
$W^\mathrm{in}_\mathbf{k}
=\sum_{\mathbf{k}^\prime}\mathcal{W}_{\mathbf{k}^\prime\rightarrow\mathbf{k}}N_{\mathbf{k}^\prime}$
are the outgoing and incoming rates for the $\mathbf{k}$-state due to scattering to and from the $\mathbf{k}^\prime$-states, respectively. The single-polariton rates $\mathcal{W}_{\mathbf{k}\rightarrow\mathbf{k}^\prime}$ account for the cumulative effect of polariton-phonon, polariton-electron and polariton-polariton interactions.

Describing the condensation dynamics we assume that at the initial moment of time the pump $P_\mathbf{k}$ is switched on, starting the competition between the states with different $\mathbf{k}$ for the particles created by the pump. The mode with $\mathbf{k}=\mathbf{k}_c$, which wins this competition,  accumulates a macroscopically large number of polaritons, thereby manifesting the formation of the condensate. In what follows we are aimed to determine the growth rates of the polariton modes, which govern the condensate formation.

First, we neglect the transitions of particles from the lower energy states to the reservoir, since these transitions are unlikely at low temperatures. Thus the outcoming rate $W^\mathrm{out}_\mathbf{k}$ for the $\mathbf{k}$-state corresponds to the downward transitions only and
$W^\mathrm{out}_\mathbf{k}=\sum_{|\mathbf{k}^\prime|<|\mathbf{k}|} \mathcal{W}_{\mathbf{k}\rightarrow\mathbf{k}^\prime}
\left(N_{\mathbf{k}^\prime}+1\right)$.
In the parabolic region of the polariton dispersion one can approximate $W^\mathrm{out}_{\mathbf{k}}\simeq\gamma k^2$, where the factor $\gamma$ is defined by the relaxation mechanism \cite{Cristofolini2013,Wouters2010}.

Next, we note that the pump $P_\mathbf{k}$ feeds mainly the exciton-like large wave vector states. These excitons form an incoherent reservoir with a particle concentration $N_r$ and they are responsible for the pumping of the lower energy polariton states with the rate $W^\mathrm{in}_{\mathbf{k}}$. With the same accuracy, the $\mathbf{k}$-dependence of the incoming rate is assumed to be parabolic. For the static in average reservoir, it is natural to assume that the $\mathbf{k}=0$ state has the maximum gain and $W^\mathrm{in}_{\mathbf{k}}\propto(1-sk^2)$. Then, if the reservoir is in motion characterized by the mean wave vector $\mathbf{k}_r$, we can write
\begin{equation}\label{Eq.IncomingRate}
  W^\mathrm{in}_{\mathbf{k}}=r(\mathbf{k})N_r, \qquad
  r(\mathbf{k}) \simeq r_0\left[1-s(\mathbf{k}-\mathbf{k}_r)^2\right].
\end{equation}
The parameter $s$ in this expression can be estimated by assuming that  excitons in the reservoir obey the Boltzmann statistics. We shall also assume that the wave vector dependence of the scattering probability is governed by the characteristic scattering cross-section dependent on the thermal De Broglie wavelength of the scatterer  $\lambda_\mathrm{th}=\sqrt{\frac{2\pi\hbar^2}{m_\mathrm{ex} k_B T}}$, where $k_B$ is the Boltzmann constant \cite{Huang1997}. We estimate $s=\lambda_\mathrm{th}^2\simeq0.058$~$\mu$m$^2$ for the exciton mass $m_\mathrm{ex} \sim 0.1m_0$ and at the temperature $T\sim 1\;$K.

In order to account for the superfluid dynamics we supplement the rate equations \eqref{Eq:Kinetic} with the generalised Gross-Pitaevskii equation for the condensate wave function $\Psi$, similar to \cite{Wouters2007}, which in our case takes into account the $\mathbf{k}$-dependencies of the gain and the loss rates:
\begin{subequations}\label{GPE+RES}
\begin{align}
\partial_t \Psi &= \hat{\mathcal{D}}\Psi - i \hbar^{-1} \hat{\mathcal{H}} \Psi, \label{GPE}\\
\partial_t N_r &= P - \left[\Gamma_r + R(\Psi)\right]N_r - \bm{\nabla}\cdot\mathbf{J}_r. \label{ResEq}
\end{align}
\end{subequations}
Here $\hat{\mathcal{H}} = -\frac{\hbar^2}{2m_{\rm pol}}\nabla^2 + \alpha_c \left|\Psi\right|^2 + \alpha_r N_r $, with $m_{\rm pol}$ being the polariton mass and the $\alpha_c$ and $\alpha_r$ terms describing the energy shifts due to interaction of polaritons between themselves and with reservoir excitons, respectively.
The gain-dissipation term with $\hat{\mathcal{D}}= 1/2\ (N_r\hat{r} - \hat{\Gamma}  )$ is given by the gain operator $\hat{r}\equiv r(-i\bm{\nabla})$ and the dissipation operator $\hat{\Gamma} = \Gamma_0 - \gamma\nabla^2$, where $\Gamma_0$ is the radiative loss rate, which is taken identical for all the states at the bottom of the lower polariton branch.
In   Eq.~\eqref{ResEq}, the term $P$ accounts for the effective pumping of the reservoir, $\Gamma_r$ describes the reservoir decay rate. We impose the local conservation of the number of polaritons in the process of their relaxation to the condensate, so that the corresponding reservoir depletion rate is given by $R(\Psi) = 1/2\left(\Psi^* \hat{r} \Psi + \Psi \hat{r}^\dag \Psi^* \right)$.

\begin{figure*}
\includegraphics[width=\linewidth]{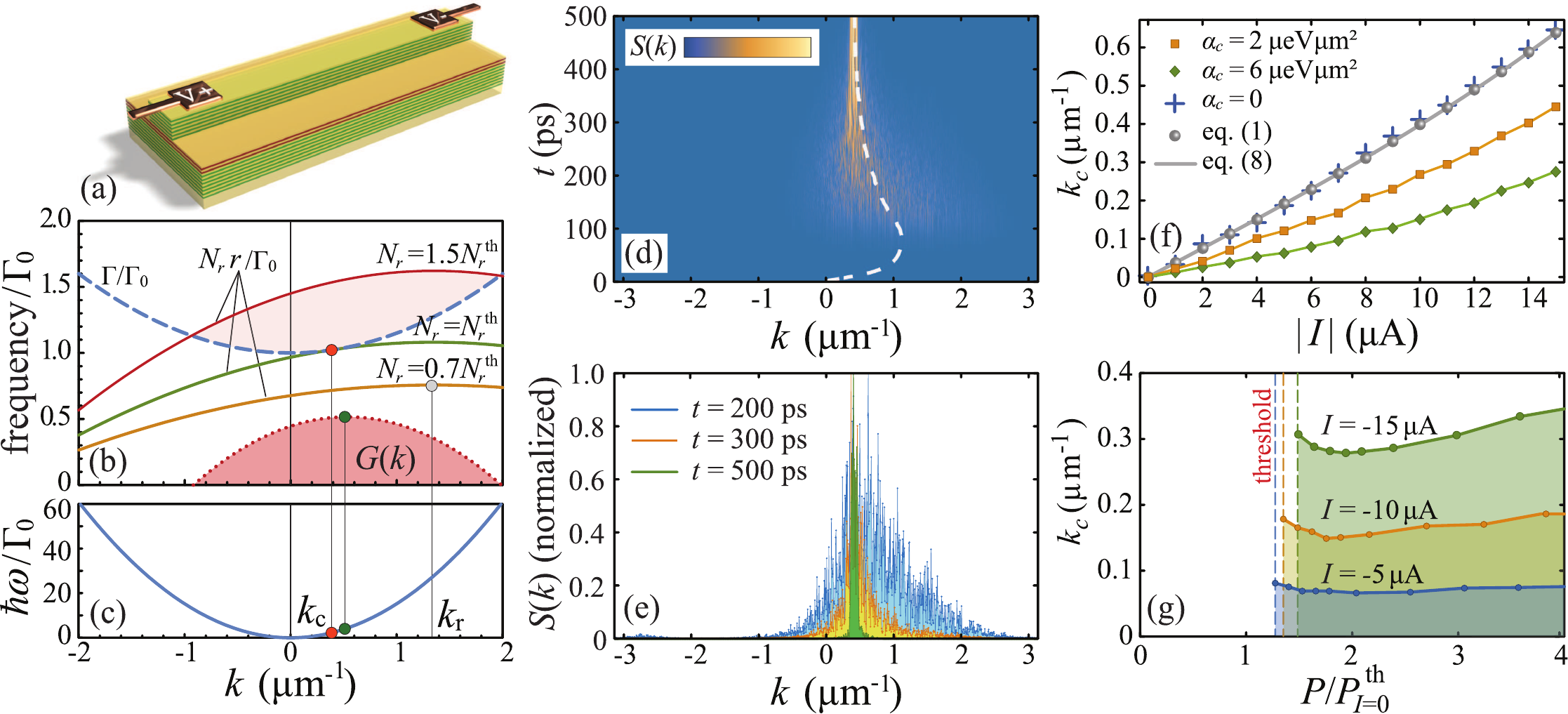}
\caption{The pseudo-drag effect in an electrically biased microcavity stripe. (a)  The sketch of the structure.  (b)  The distribution of the gain $N_r r(\mathbf{k})$ (solid) and the loss rate $\Gamma(\mathbf{k})$ (dashed)  in the reciprocal space at the reservoir densities below threshold (orange curve), at the threshold (green) and above threshold (red). The mean wave vector of the reservoir is $k_r=1.35$~$\mu$m$^{-1}$.
The shaded domain shows the net gain $G(\mathbf{k})$ at the pump power corresponding to the red curve. (c)  The polariton dispersion.
(d) The dynamics of the condensate formation in the reciprocal space. $P=2P^{\rm th}_{I=0}$, $\alpha_c=6$~$\mu$eV$\mu$m$^2$, $k_r=1.35$~$\mu$m$^{-1}$, where $P^{\rm th}_{I=0}=\Gamma_0\Gamma_r/r_0$. White dashed line follows the position of the mean condensate wave vector. (e) The condensate spatial spectra at three different moments of time. For each curve the spectrum maximum is normalized to unity.
(f) The dependence of the condensate wave vector $k_{\rm c}$ on the electric current flowing though the stripe of $d=4$~$\mu$m width. Blue crosses, orange squares and green diamonds show the data obtained with Eqs.~\eqref{GPE+RES}. The dynamics was simulated with noisy initial conditions and the data were averaged over 50 realizations for each point. Grey points indicate the predictions of Eqs.~\eqref{Eq:Kinetic}. Grey line corresponds to Eq.~\eqref{Eq.kc}. (g) The dependence of $k_{\rm c}$ on the  pump power $P$ for different values of the electric current $I$ with $\alpha_c=6$~$\mu$eV$\mu$m$^2$.
For all panels $\Gamma_0=0.05$~ps$^{-1}$, $\gamma=7.5\times 10^{-3}$~$\mu$m$^2$~ps$^{-1}$, $r_0=0.01$~ps$^{-1} \mu$m$^2$, $\Gamma_r=5\Gamma_0$, $\alpha_r=2\alpha_c$.} \label{Fig.f1}
\end{figure*}

The last term in Eq.~\eqref{ResEq} provides the continuity of the mass flow in the presence of a spatially inhomogeneous current $\mathbf{J}_r =  \hbar \mathbf{k}_r N_r \left/ m_{\rm ex} \right.$ of reservoir excitons. The reservoir flux can be created either by the exciton density gradient or by the external force, in particular, by the Coulomb drag. The value of the exciton current is dependent on the  exciton and carrier densities as well as on their relative velocity. Namely,
\begin{equation}\label{Eq.exciton_current}
\mathbf{J}_r =  -D \bm{\nabla} N_r + \lambda\left(\mathbf{v} - \frac{ \hbar\mathbf{k}_r}{m_{\rm ex}} \right) n N_r,
\end{equation}
where $D$ is the exciton diffusion coefficient, $\lambda$ is the drag coefficient, $\mathbf{v}$ and $n$ are the drift velocity and the surface density of the carriers.

The drag coefficient $\lambda$ can be estimated from Ref.~\cite{berman_10}, to be $\lambda \approx 1.1\times 10^{-12}$~cm$^2$ for the case of the conducting layer represented by  the n-doped GaAs quantum well with $n=10^{11}$~cm$^{-2}$  and  the electron mobility \cite{Basu1991}
$\mu=10^3$~cm$^2$V$^{-1}$s$^{-1}$.

\section{Polariton pseudo-drag effect in a microcavity stripe}\label{secIII}

A remarkable manifestation of the pseudo-drag effect could be found in the double layer system embedded in a microcavity stripe, see Fig.~\ref{Fig.f1}(a). We shall consider a  one-dimensional  exciton-polariton condensate   excited by the homogeneous nonresonant  continuous wave laser pumping.

The voltage applied at the metal contacts \cite{Snoke2018} induces the electric current in the conducting layer, which drags the reservoir excitons in the second quantum well with the mean wave vector
\begin{equation}\label{Eq.ResWV}
\mathbf{k}_r=-\xi \mathbf{j},
\end{equation}
defined by Eq.~\eqref{Eq.exciton_current} with $\mathbf{j}=-en\mathbf{v}$ being the electric current density. Here the factor $\xi={m_{\rm ex} \lambda }\left/\left({\hbar e\left(1+\lambda n\right)}\right)\right.$ for the considered parameters equals to $ 5.4 \times 10^{5}$~A$^{-1}$. It implies that for  $d=4\;\mu$m  stripe width the reservoir flow with ${k}_r=1\ \mu{\rm m}^{-1}$ can be induced by an electric current of $I=j d\approx -7.4$~$\mu{\rm A}$. This value agrees with the recent experimental studies of the drag effect in a polariton wire structure  \cite{Snoke2018}.

\begin{figure*}
\includegraphics[width=0.9\linewidth]{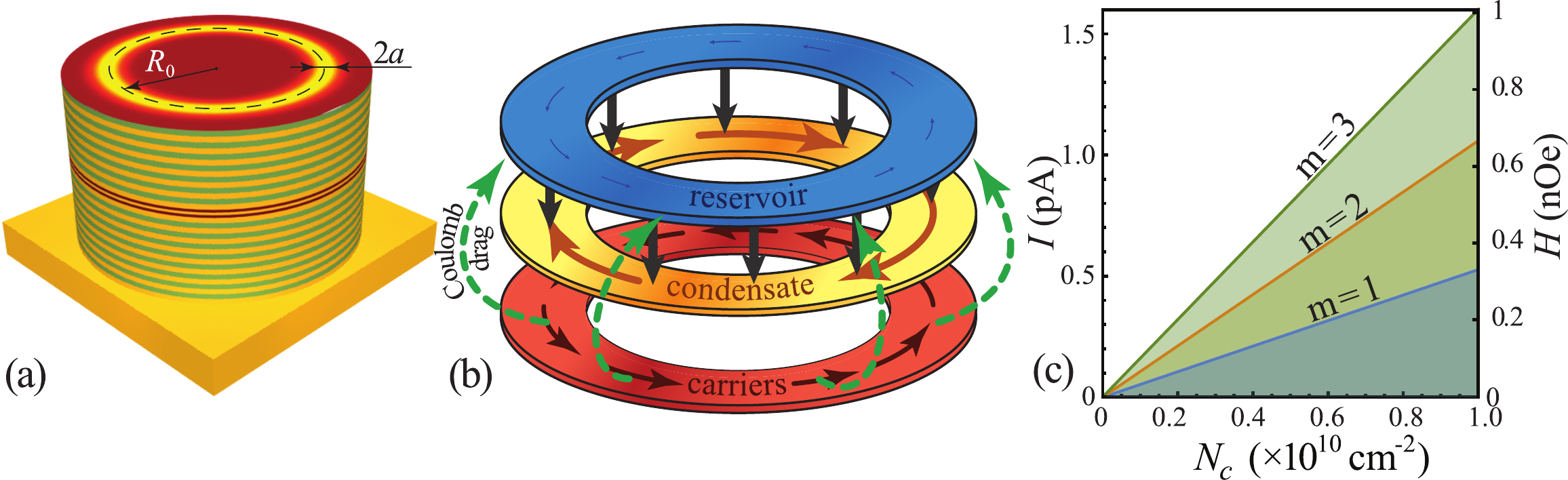}
\caption{Generation of the circular electric current supported by the electron-mediated exciton scattering. (a) A pillar microcavity with the embedded two-layer system. The condensate density $N_c$ distribution is sketched at the top surface of the micropillar. (b) The schematic illustration  of the electric current generation mechanism. (c) The dependence of the  electric current $I$ and the corresponding magnetic field $H$ at the center of the quantum well plane on the  condensate density $N_c$.}\label{Fig.f2}
\end{figure*}

The condensate steady state  reads $\Psi=\Psi_0 e^{-i\omega t + ik_c x}$, where $x$ is the coordinate along the stripe and the wave vector $\mathbf{k}_c$ is selected during the condensate formation governed by the $\mathbf{k}$-dependent growth rate
\begin{equation}\label{Eq.GrowthRate}
G(\mathbf{k},t)=2\Psi_0^{-1} \hat{\mathcal{D}}\Psi= N_r(t)r(\mathbf{k}) - \Gamma (\mathbf{k}),
\end{equation}
where $\Gamma(\mathbf{k})=\Gamma_0+\gamma k^2$.

The principles of the $\mathbf{k}$-selective mechanism of the condensate formation are described below. As the pump $P$ switches on, the reservoir population starts growing. The condensate builds up provided that the reservoir density reaches the
threshold value (see the green curve and the red points in Figs.~\ref{Fig.f1}(b),(c)):
\begin{equation}\label{Eq.N_res^th}
N^{\rm th}_{r}=\frac{\Lambda+\sqrt{\Lambda^2 + 4\gamma\Gamma_0s}}{2r_0s},
\end{equation}
where $\Lambda=\Gamma_0s-\gamma\left(1-sk^2_r\right)$.
Above the threshold, $P>\Gamma_r N^{\rm th}_{r}$, a wide band of polariton states (the shaded region in Fig.~\ref{Fig.f1}(b)) is amplified at the initial stage of the condensate formation -- see Figs.~\ref{Fig.f1}(d),(e).

However, as the condensate population increases, the reservoir becomes depleted because of the term $R\left(\Psi\right)$ in Eq.~\eqref{ResEq}. Simultaneously, the amplification band maximum shifts towards smaller $\mathbf{k}$ resulting in the drift of polaritons in the reciprocal space -- Fig.~\ref{Fig.f1}(e). The analysis of the condensation  dynamics predicted by Eqs.~\eqref{GPE+RES} (see Appendix) demonstrates that in the limit of no interactions ($\alpha_c=0$) the reservoir eventually relaxes to its threshold value  \eqref{Eq.N_res^th}, which corresponds to  the condensate state with
\begin{equation}\label{Eq.kc}
\mathbf{k}_{c}=\frac{sr_0 N^{\rm th}_r \mathbf{k}_r}{sr_0 N^{\rm th}_r + \gamma}.
\end{equation}
Therefore, the condensate momentum shown in Fig.~\ref{Fig.f1}(f) grows with the electric current since $\mathbf{k}_r \propto I$, see Eq.~\eqref{Eq.ResWV}.
The same result can be obtained from the kinetic equations \eqref{Eq:Kinetic} as well (see Appendix). Besides, Eq.~\eqref{Eq.kc} is valid not only for the case of microcavity stripe but for 2D polariton condensates as well.
Note that the energy relaxation of polaritons is taken into account in the present theory by the parameter $\gamma$, which encodes the increase of the outcoming (escape) rate of polaritons with increasing momentum. This leads to effective friction, so that $\mathbf{k}_{c}$ decreases with $\gamma$ (see Appendix).

In the nonlinear regime, the  parametric scattering between polariton modes caused by polariton-polariton interactions leads to the reduction of  the condensate momentum, as Fig.~1(f) illustrates. The higher the value of the interaction strength $\alpha_c$, the stronger the decrease of the condensate momentum with respect to the limiting value predicted by Eq.~\eqref{Eq.kc}, see the orange squares and green diamonds in Fig.~\ref{Fig.f1}(f). The condensate momentum demonstrates also a weak dependence on the pump power, see Fig.~\ref{Fig.f1}(g).

\section{Polariton whirl producing a circular electric current}

So far, we neglected   the back action of the exciton-polariton subsystem on the electron gas. This action is twofold. First,  due to the inverse Coulomb drag effect \cite{berman_10,Narozhny2016} the exciton flow induces an electric current in the conducting layer. This effect can be accounted for in a similar way to Eq.~\eqref{Eq.exciton_current}:
\begin{equation}
\mathbf{j}_{\rm drag}=-\nu \left(\frac{\hbar\mathbf{k}_r}{m_{\rm ex}} - \mathbf{v}\right) n N_r,
\end{equation}
where $\nu$ is the drag coefficient and we consider again the n-doped conducting layer.

Besides this, there is the stimulated electron-assisted scattering of excitons from the reservoir to the condensate. Each act of electron-assisted scattering transfers the momentum $\hbar\delta\mathbf{k}=\hbar(\mathbf{k}_r-\mathbf{k}_c)$ to the electron gas. This momentum transfer can be described in terms of a force acting on a single electron:
\begin{equation}
\mathbf{F}=\beta\frac{\hbar N_r R(\Psi)}{n}\delta\mathbf{k},
\end{equation}
where $N_r R(\Psi)$ is the scattering rate defined in Eq.~\eqref{ResEq} and the factor $\beta<1$ accounts for the part of the total number of electron-mediated scattering events. Then, the current $\mathbf{j}_{\rm scat}$ induced by the electron-mediated scattering can be calculated using the classical Drude theory of conductivity:
\begin{equation}\label{Eq.ScatCurr}
\mathbf{j}_{\rm scat}=  \sigma  \delta\mathbf{k} N_r R(\Psi),
\end{equation}
where $\sigma=\hbar\beta\mu$, which for the considered parameters and $\beta=0.05$ is $\sigma=5.26\times 10^{-25}\,\mathrm{C}\,\mathrm{cm}^2 $.

Note, that both discussed mechanisms are capable of inducing an electric current at zero voltage. However, in contrast to the Coulomb drag effect, the electron-mediated scattering induces the current which is dependent on the state of polariton superfluid. It allows for nontrivial manifestations of the effect. For instance, a spectacular property of the exciton-polariton condensates is its ability to form a persistent circular current or vorticity. Once such a circular current is created, its non-zero angular momentum is partially transferred to the electron gas producing a circular current as Fig.~\ref{Fig.f2}(b) schematically shows.

Let us study these phenomenon in detail. Although polariton condensates carrying   non-zero angular momenta were obtained in many different configurations \cite{LagoudakisK2008,Roumpos2011}, to be specific we focus on the system shown in  Fig.~\ref{Fig.f2}(a). We shall assume that the double layer system  is embedded in a cylindrical microcavity pillar, where the formation of a ring-shaped condensate carrying the persistent polariton current was recently demonstrated \cite{Lukoshkin2018}.

The distribution of the condensate density $N_c=\left|\Psi\right|^2$ is characterized by the mean radius $R_0$ and the width $2a$, see Fig.~\ref{Fig.f2}(a). Then, assuming that $N_c$ takes the constant value over the width of the ring \cite{li2015}, we estimate the rate of scattering as $N_r R(\Psi) = r(\mathbf{k}) N_c N_r^{{\rm st}}$, where the condensate wave vector is defined by the winding number $m$ as $k_c=m/R_0$. The reservoir steady-state density $N_r^{{\rm st}}$ can be obtained from Eq.~\eqref{ResEq}:
\begin{equation}\label{Eq.Nrst}
N_r^{{\rm st}} = \frac{\Gamma_0 + \gamma k_c^2}{ r_0\left( 1 - s (\mathbf{k}_c-\mathbf{k}_r)^2\right)}.
\end{equation}

In order to estimate the value of the induced current $\mathbf{I}_{\rm scat}= 2a\mathbf{j}_{\rm scat}$, we assume that the reservoir is at rest, $\mathbf{k}_r=0$, and take $m=1$, $a=1$~$\mu$m and $R_0=10$~$\mu$m. So, for the typical value of the condensate density $N_c=10^{10}$~cm$^{-2}$ we obtain that the current is on a picoampere scale. According to Eq.~\eqref{Eq.ResWV} this current induces  a negligibly small flow of reservoir excitons (see Fig.~\ref{Fig.f2}(b)), that is consistent with our assumption of the static  reservoir.

The existence of the predicted circular electric current can be experimentally detected measuring the current-induced magnetic field, for instance, with state of the art SQUID magnetometers \cite{tumanski2016}. Assuming that the circular current density $\mathbf{j}_{\rm scat}$ follows the condensate density $N_c$ distribution, we estimate the magnetic field in the center of the ring, $H=I_\mathrm{scat}/2R_0$.  The dependence of the induced electric current given by Eqs.~\eqref{Eq.ScatCurr}, \eqref{Eq.Nrst} and the corresponding magnetic field on the condensate density is shown in Fig.~\ref{Fig.f2}(c).

\section{Conclusions}

We demonstrate that propagation of out-of-equilibrium polariton condensate can be controlled by an electric current, as the latter influences the gain rate of different single-polariton quantum states. The effect appears due to drift of the excitonic reservoir. The existence of this pseudo-drag effect paves the way for the engineering of the integrated optical circuits operating with polariton condensates. It may be promising for the creation of superfluid gyroscopes and quantum interferometers. The reciprocal effect of acceleration of the charge carriers by the moving polariton superfluid is also described.  Being sustained by the stimulated exciton-electron scattering, the carrier's flow is evidenced by the circular electric current in a cylindrical micropillar excited by a nonresonant laser pump.

\begin{acknowledgments}
We thank D. Snoke and A. Varlamov for many enlightening discussions.
This work is supported by Westlake University (Project No.\ 041020100118). IYC acknowledges support from RFBR under grant No.\ 17-52-10006 and from the Ministry of Science
and Higher 	Education of the Russian Federation (state project No.\ 16.1123.2017/4.6). YGR acknowledges support from CONACYT (Mexico) under Grant No.\ 251808.
\end{acknowledgments}

\appendix
\section{Dynamics of the condensate formation in the presence of the momentum-dependent gain}\label{appendix}

Here we describe in details how the presence of the momentum-dependent gain affects the polariton condensation dynamics in the microcavity stripe leading to the formation of the moving condensate. We start with the kinetic equations \eqref{Eq:Kinetic} for polariton modes populations, whose simplified form reads:
\begin{subequations}\label{Eq.Kinetic}
\begin{eqnarray}
\partial_t {N}_\mathbf{k} &=& r(\mathbf{k})\left(N_\mathbf{k}+1\right)N_r - \Gamma(\mathbf{k})N_\mathbf{k}, \label{Eq.Kinteic.Nk}\\
\partial_t {N}_r&=&P- \left(\Gamma_r + \sum_\mathbf{k}{r(\mathbf{k})\left(N_\mathbf{k}+1\right)}\right)N_r. \label{Eq.Kinteic.Nr}
\end{eqnarray}
\end{subequations}
Here  $N_r$ is the spatially homogeneous reservoir density, while $N_\mathbf{k}$ denotes the total number of polaritons in the quantum state $\mathbf{k}$. 

\begin{figure}
\includegraphics[width=\linewidth]{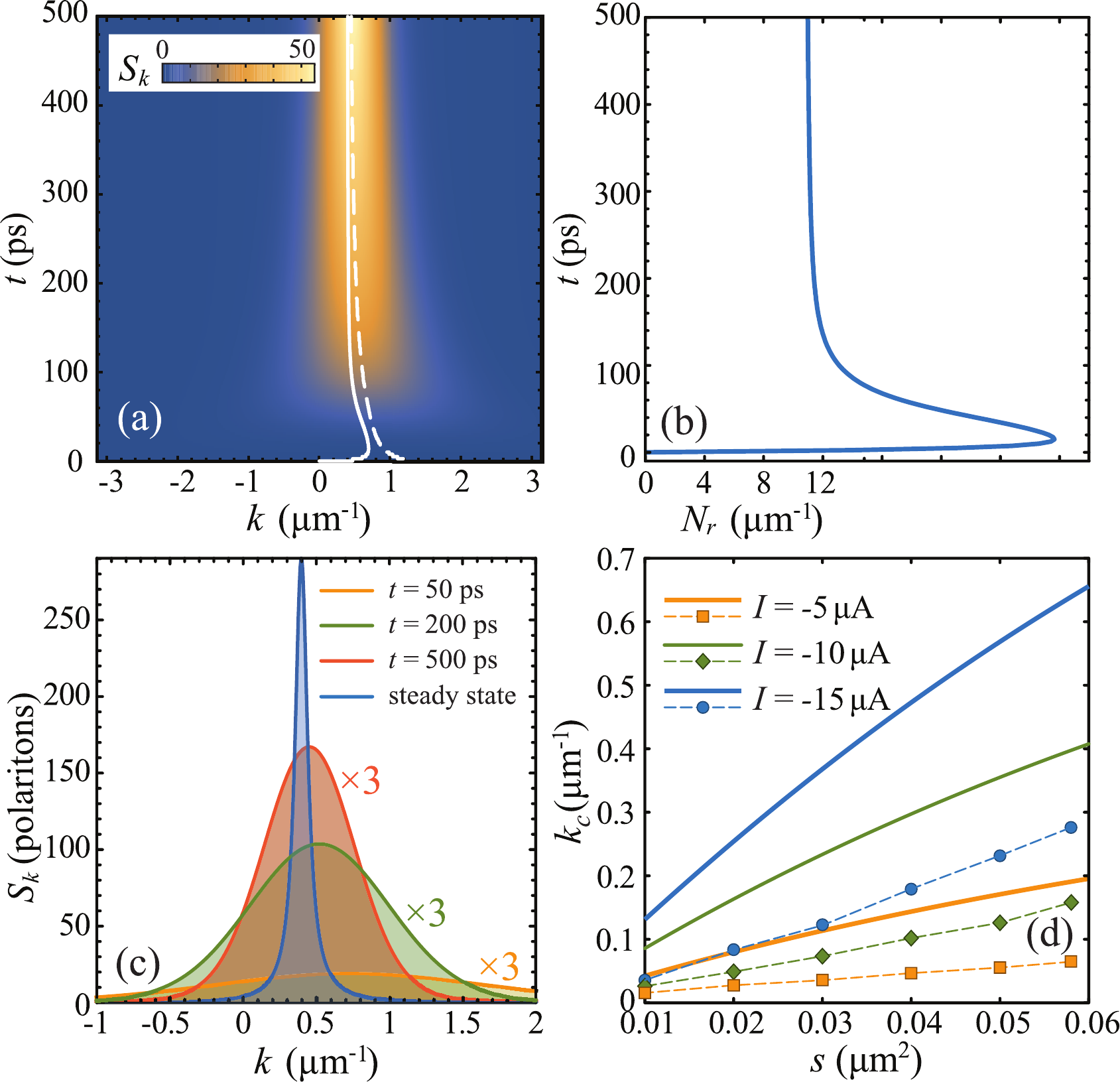}
\caption{The polariton condensate formation in the microcavity stripe. 
(a) Dynamics of the polariton distribution in the reciprocal space predicted by the kinetic equations model~\eqref{Eq.Kinetic}. Colorscale corresponds to the spectral density of polaritons per micrometer, $S_\mathbf{k} \equiv N_\mathbf{k}\left/ 2\pi \right.$. The white solid line shows the position of the gain maximum $\mathbf{k}_{{\rm max}(G)}$. The dashed line indicates the maximum of the polariton spectral density. (b) Dynamics of the reservoir density. (c) Snapshots of the polariton distibution at different moments of time. The blue curve corresponds to the steady-state distribution. The spacing between the polariton states in the reciprocal space is $\pi/500$~$\mu$m$^{-1}$. (d) The dependence of the condensate wave vector $k_c$ on the $s$-parameter for different values of the electric current. The solid curves correspond to Eq.~\eqref{Eq.kinetic.kc}, while the markers indicates the values predicted by the generalized Gross-Pitaevskii model with $\alpha_c=6$~$\mu$eV$\mu$m$^2$ and $\alpha_r=2\alpha_c$.
The parameters are $k_r=1.35\ \mu{\rm m}^{-1}$, $P=3\Gamma_0\Gamma_r/r_0$, $\Gamma_0=0.05$~ps$^{-1}$, $\gamma=7.5\times 10^{-3}$~$\mu$m$^2$~ps$^{-1}$, $r_0=0.01$~ps$^{-1} \mu$m$^2$, $\Gamma_r=5\Gamma_0$.}\label{Fig.Sf1}
\end{figure}

The dynamics predicted by Eqs.~\eqref{Eq.Kinetic} is illustrated in Fig.~\ref{Fig.Sf1}. We assume that before the pump has been switched on at $t=0$, the polariton modes are empty, $N_k=0$. Then, at the initial stage of the condensate formation, the growth of the population of polariton modes is governed by the term $r(\mathbf{k})N_r$, see Eq.~\eqref{Eq.Kinteic.Nk}, which follows the $\mathbf{k}$-dependence of the rate of the income transitions  $r(\mathbf{k})$  and is maximized at $\mathbf{k}=\mathbf{k}_r$, see the yellow curve in Fig.~\ref{Fig.Sf1}(c). However,  the gain profile changes as the polariton modes get populated. In particular, at $N_\mathbf{k} \gg 1$, the dynamics obeys to the net gain, which maximum is at
\begin{equation}
\mathbf{k}_{{\rm max}(G)}= \frac{sr_0N_r(t)}{sr_0N_r(t)+\gamma}\mathbf{k}_r.
\end{equation}
As far as  $k_{{\rm max}(G)} < k_r$, the gain $G(\mathbf{k})$ causes redistribution of  population between the polariton modes excited before   shifting its maximum towards smaller $\mathbf{k}$, see Fig.~\ref{Fig.Sf1}(c).

It is critical that the net gain $G(\mathbf{k},t)$ is time-dependent since it is governed by the  reservoir density $N_r$. As the pump power switches instantly, the reservoir rapidly grows approaching its trivial steady state $N_r=P/\Gamma_r$, see Fig.~\ref{Fig.Sf1}(b). However, after the burst at the initial stage, the condensate density smoothly decreases towards its stationary value $N_r=N_r^{\rm st}$, where the dissipation and the out-scattering to the condensate balances the pump. The position of the gain maximum $\mathbf{k}_{{\rm max}(G)}$, indicated by  the white dashed line in Fig.~\ref{Fig.Sf1}(a), follows the reservoir density evolution. Simultaneously, the polariton distribution approaches the position of the maximum gain gradually evolving to its steady state (see the white solid  line in Fig.~\ref{Fig.Sf1}(a)):
\begin{equation}\label{Eq.kinetic.StedySt}
N_\mathbf{k}=\frac{r(\mathbf{k})N_r^{\rm st}}{\Gamma(\mathbf{k}) - r(\mathbf{k})N_r^{\rm st}}.
\end{equation}

Note, that the Botzmann equations model predicts that in the steady state the condensate occupies a narrow band of states in the reciprocal space. The shape of this band is defined by Eq.~\eqref{Eq.kinetic.StedySt}, see the blue line in Fig.~\ref{Fig.Sf1}(c). The position of the maximum of polariton distribution, which can be associated with the condensate wave vector,
\begin{equation}\label{Eq.kinetic.kc}
k_c=\frac{\Lambda - 2s \Gamma_0 + \sqrt{4k_r^2s^2\gamma \Gamma_0 + \left(\Lambda- 2s \Gamma_0\right)^2}}{2k_rs\gamma},
\end{equation}
precisely matches the expression \eqref{Eq.kc}, which was obtained with use of the generalized Gross-Pitaevskii model neglecting polariton-polariton interactions. 
Note that the condensate momentum does not depend on the pumping power $P$ and is governed by the parameters of the $\mathbf{k}$-dependent gain. In Fig.~\ref{Fig.Sf1}(d) we demonstrate the variation of the steady-state condensate momentum with the variation of the effective
scattering cross-section described by the parameter $s$, which governs the $\mathbf{k}$-dependence of the incoming rate.

The  kinetic equations model clearly demonstrates the influence of the reservoir dynamics on  the reciprocal space polariton distribution. However, the considered approach neglects polariton-polariton interactions as well as the spatial dependence of the reservoir density. Both these effects can be efficiently  accounted with the driven-dissipative Gross-Pitaevskii (GP) model \eqref{GPE+RES}.

An example of the condensate formation dynamics predicted by the generalized GP model is shown in Figs.~\ref{Fig.f1}(d),(e). The overall condensation scenario discussed in Sec.~\ref{secIII} is akin to the one described above. The mean condensate wave vector follows the time dependence of the net gain $G(k)$ which is governed by the reservoir population. After a rapid grows at the initial stage, the wave vector relaxes to the steady state (the dashed curve in Figs.~\ref{Fig.f1}(d)). 
However, in contrast to the case of kinetic equations~\eqref{Eq.Kinetic}, the generalized GP model has a set of steady state solutions  parameterized by the condensate wave vector $k_c$. Namely,
\begin{equation}
\Psi= \Psi_0 e^{ik_c x - i\omega t},\ \ \ \  N_r^{\rm st} =\frac{ \Gamma(k_c)}{r(k_c)},
\end{equation}
with $\left| \Psi_0\right|^2={P}\left/{\Gamma(k_c)}\right. - {\Gamma_r}\left/{r(k_c)}\right.$. It implies that the condensate can exist in an arbitrary $\mathbf{k}$-state within the amplification band, $G\left(N_r=P/\Gamma_r\right)>0$.

The particular state  of the condensate, selected during its formation, is strongly dependent on the initial conditions. In stark contrast to the model \eqref{Eq.Kinetic} the GP equation requires the initial seed in order to produce the nontrivial steady state. We simulate the condensate formation with the model \eqref{GPE+RES} using random initial conditions, which mimic the thermal fluctuations of the polariton field. It is clear that the choice of the condensate wave vector is stochastic in this case. In the linear case ($\alpha_c=\alpha_r=0$), the  averaging over large number of realizations yields the value $k_c$ which coincides with the one predicted by Eq.~\eqref{Eq.kinetic.kc}. One can easily demonstrate  that this state corresponds to the threshold value of the reservoir density $N_r=N_r^{\rm {th}}$ defined by Eq.~\eqref{Eq.N_res^th}.

\begin{figure}
\includegraphics[width=\linewidth]{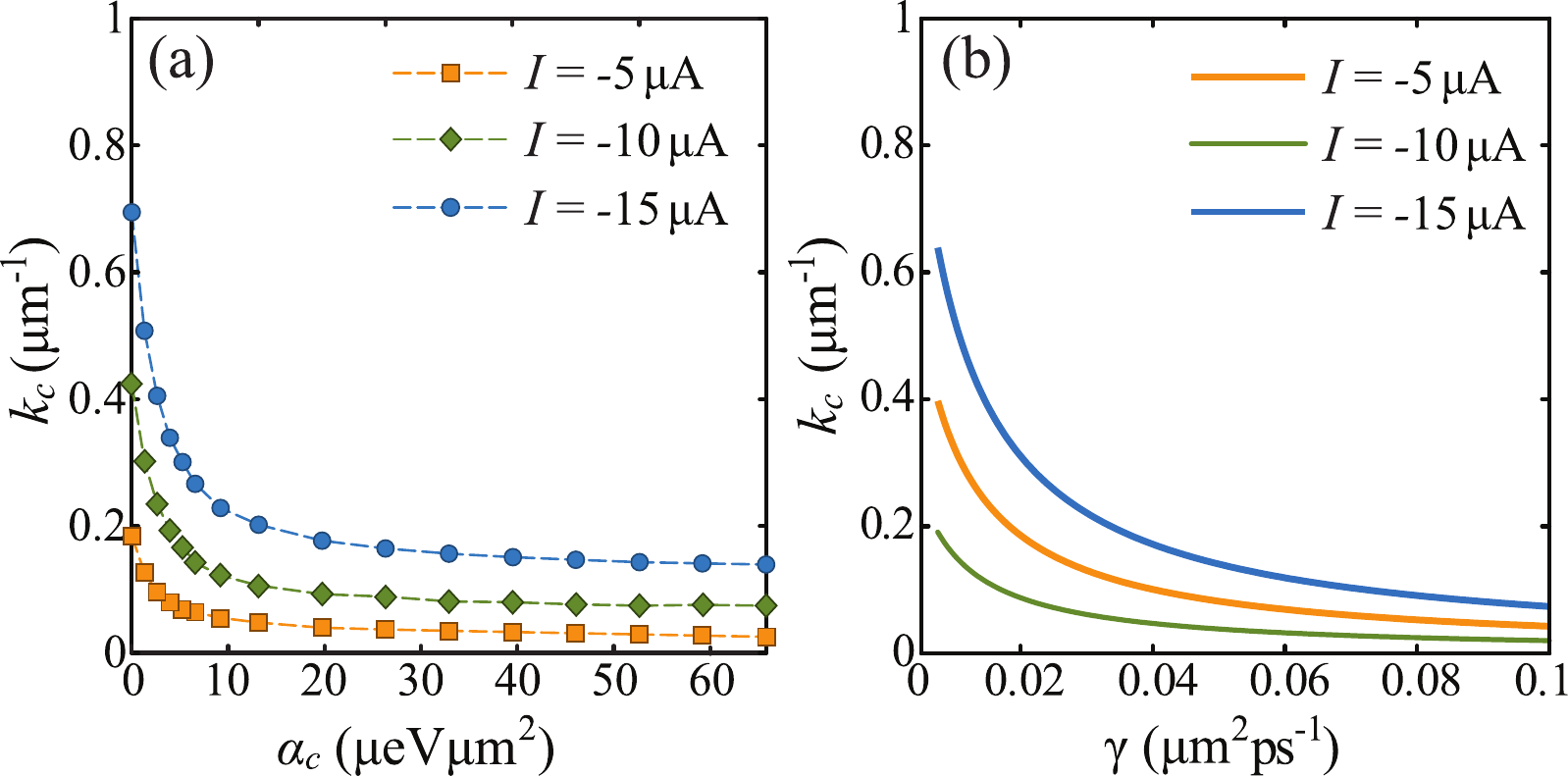}
\caption{(a) The dependence of the pseudo-drag induced momentum of the polariton superfluid on the polariton-polariton interaction constant $\alpha_c$. The polariton-reservoir interaction strength is $\alpha_r=2\alpha_c$. (b) The dependence \eqref{Eq.kc} of the condensate wave vector $k_c$ on the $\gamma$-parameter for different values of the electric current. }\label{Fig.Sf2}
\end{figure}

In the presence of polariton-polariton and polariton-reservoir interactions, the dynamics of the condensate formation follows the same scenario, however the average condensate wave vector is reduced, see color markers in Fig.~\ref{Fig.Sf1}(d) and compare with solid curves corresponding to the linear case. The influence of the polariton-polariton interaction constant $\alpha_c$ on the condensate momentum is summarized in Fig.~\ref{Fig.Sf2}. For the polariton nonlinearities typical for GaAs-based microcavities ($\alpha_c$ is several units of $\mu\rm{eV}\mu\rm{m}^2$), the condensate momentum steeply decreases with the increase of $\alpha_c$, see Fig.~\ref{Fig.Sf2}(a). Then, $k_c$ gradually reduces to zero as $\alpha_c$ grows.

The observed reduction of the pseudo-drag effect in the presence of interactions should be attributed to the redistribution of polaritons in the reciprocal space induced by the polariton-polariton repulsion. Indeed, spatial inhomogeneities of the polariton density, grown from the thermal noise during the condensate formation, are smoothed away because of the polariton repulsion. The excited high-momentum polaritons are subject to the extra losses accounted in the $\Gamma(k)$-dependence, that is $\Gamma(k)=\Gamma_0+\gamma k^2$ in the 1D case, see Eq.~\eqref{GPE+RES}.
Thus the influence of the interactions can be accounted by rescaling of the $\gamma$-parameter. 
According to Eq.~\eqref{Eq.GrowthRate}, an increase of the steepness of the $\Gamma(k)$-dependence shifts an amplification band towards smaller polariton momenta reducing the wave vector of the condensate, see Fig.~\ref{Fig.f1}(b).  Thus, the condensate wave vector predicted by Eq.~\eqref{Eq.kc}   should decrease as it shown in Fig.~\ref{Fig.Sf2}(b). This behaviour qualitatively reproduces the $k_c(\alpha_c)$-dependence shown in Fig.~\ref{Fig.Sf2}(a).

%

\end{document}